\documentclass[a4paper]{revtex4}
\usepackage{amsfonts}
\usepackage{amsmath}
\usepackage{amssymb}
\usepackage{graphicx}
\usepackage{dsfont}

\newtheorem{prop}{PROPOSITION}[section]

\begin{document}

\title{Characterization of fiducial states in prime dimensions\\ via mutually unbiased bases\vspace{0.3cm}}
\author{D. Goyeneche}
\email{dardo.goyeneche@cefop.udec.cl}
\author{R. Salazar}
\author{A. Delgado}

\affiliation{\vspace{0.5cm}Departamento de Fis\'{i}ca, Universidad de Concepci\'{o}n, Casilla 160-C, Concepci\'{o}n, Chile\\Center for Optics and Photonics, Universidad de Concepci\'{o}n, Casilla 4016, Concepci\'{o}n, Chile\vspace{0.5cm}}

\begin{abstract}
In this work we present some new properties of fiducial states in prime dimensions. We parameterize fiducial operators on eigenvectors bases of displacement operators, which allows us to find a manifold $\Omega$ of hermitian operators satisfying $\mathrm{Tr}(\rho)=\mathrm{Tr}(\rho^2)=1$ for any $\rho$ in $\Omega$. This manifold contains the complete set of fiducial pure states in every prime dimension. Indeed, any quantum state $\rho\geq0$ belonging to $\Omega$ is a fiducial pure state. Also, we present an upper bound for every probability associated to mutually unbiased decomposition of fiducial states. This bound allows us to prove that every fiducial state tends to be mutually unbiased to the maximal set of mutually unbiased bases in higher prime dimensions. Finally, we show that any $\rho$ in $\Omega$ minimizes an entropic uncertainty principle related to the second order R\'enyi entropy.\vspace{0.3cm}

\end{abstract}

\maketitle

\section{Introduction}
A symmetric informationally complete positive-operator-valued measure (SIC-POVM) is a set of $d^2$ rank-one projectors acting on a $d$-dimensional Hilbert space such that the Schmidt product among any two projectors equals $1/(d+1)$ \cite{Renes1}. The statistic of measurement outcomes associated to a SIC-POVM allows to characterize unambiguously any pure or mixed quantum state. The continued interest on this subject originates in the many important practical applications of SIC-POVMs to quantum mechanics \cite{Scott1}, quantum information \cite{Medendorp} and to classical signal processing, as well as its possible relation to a better understanding of the structure of the quantum state space \cite{Appleby0}. Currently, there exist analytical constructions in $d$ = 2, 3 \cite{Delsarte}, 4, 5 \cite{Zauner1}, 6 \cite{Grassl1}, 7 \cite{Appleby01}, 8 \cite{Grassl2} $9,\dots, 13, 15$ \cite{Grassl3}, 16 \cite{Appleby02} and 19 \cite{Appleby01}, and high precision numerical solutions in all dimensions $d\le 67$ \cite{Grassl5}.

SIC-POVMs  have been characterized as $d^2$ equiangular lines through the origin in $\mathcal{C}^d$, tight complex projective 2-designs,  maximally equiangular tight frames \cite{Grassl5}, and as simplexes in the Bloch space \cite{Salazar}. Here, our study is restricted to Weyl-Heisenberg covariant SIC-POVMs. These are constructed by applying the generators of the Weyl-Heisenberg group onto a special \emph{fiducial} pure quantum state. We decompose this fiducial state as a combination of projectors onto eigenvectors of the displacement operators. In the particular case of prime dimensions, these eigenvectors form $d+1$ mutually unbiased (MU) bases. Using Zauner's conjecture we obtain a manifold $\Omega$ determined by the complete set of \emph{fiducial operators} parameterized by variables such that any fiducial pure state leading to a Weyl-Heisenberg covariant SIC-POVM is contained within this set. We show that any $\rho$ in $\Omega$ satisfies $\mathrm{Tr}(\rho)=1$ and $\mathrm{Tr}(\rho^2)=1$ and that these operators have the same geometric properties of fiducial states. Fiducial pure states can be singled out by reaching the global maximum of the funcion $\mathrm{Tr}(\rho^3)$, with $\rho$ in $\Omega$. Thereby, SIC-POVMs are naturally described as a variational problem within the set $\Omega$. We also show that every hermitian operator $\rho$ in $\Omega$ minimizes an entropic uncertainty principle and, in particular, the fiducial states. Finally, we deduce an upper bound for each probability defining a fiducial state in the MU bases decomposition. As consequence of this upper bound, fiducial pure states tend to be mutually unbiased to the maximal set of mutually unbiased bases existing in prime dimensions.

\section{SIC-POVMs and displacement operators}
If it is possible to reconstruct every quantum state from a tomographic measurement over a set of operators forming a POVM we say that it is an informationally complete POVM (IC-POVM). The most interesting case of IC-POVM is given by the SIC-POVM, because the overlap of information and errors propagation are minimized. A SIC-POVM is a positive-operator-valued measure formed by $d^2$ linearly independent rank-one projectors $M_k$ with $k=0,\dots,d^2-1$, acting on $\mathcal{C}^d$, that satisfy the property
\begin{equation}\label{SIC_cond}
    \mathrm{Tr}(M_kM_r)=\frac{d\delta_{k,r}+1}{d+1},
\end{equation}
for every $k,r=0,\dots, d^2-1$. Considering the projecting directions we can study the equivalent problem to find $d^2$ pure quantum states $|\phi_k\rangle$ instead of $d^2$ projectors. That is, assuming that $M_k=|\phi_k\rangle\langle\phi_k|$ then the inner product between two arbitrary states must fulfill the property
\begin{equation}
    |\langle\phi_k|\phi_r\rangle|^2=\frac{d\delta_{k,r}+1}{d+1}~~~\forall~k,r=0,\dots, d^2-1.
\end{equation}
There is a strong conjecture that simplifies the way to construct SIC-POVMs. In order to introduce this conjecture let us define first the displacement operators in finite dimension, given by
\begin{equation}
D_{\mathbf{p}}=\tau^{p_1p_2}X^{p_1}Z^{p_2},
\end{equation}
where $\mathbf{p}=(p_1,p_2)\in\mathbb{Z}_d^2$ and $\tau=-e^{i\pi/d}$. The operators $X$ and $Z$ are the \emph{shift} and \emph{phase} operators, defined by
\begin{equation}
    X|k\rangle=|k+1\rangle,\hspace{0.2cm} Z|k\rangle=\omega^k|k\rangle,
\end{equation}
where $\omega=e^{2\pi i/d}$, $k=0,\dots,d-1$ and $\{|k\rangle\}$ is the canonical (computational) base. If $d=2$, the displacement operators are the Pauli matrices plus the identity. These operators form, up to a multiplicative constant factor, the generalized Pauli group or Weyl-Heisenberg group. Their commutation rule is given by
\begin{equation}\label{Conmutation}
 D_{\mathbf{p}}D_{\mathbf{q}}=\tau^{\langle \mathbf{p},\mathbf{q}\rangle-\langle \mathbf{q},\mathbf{p}\rangle}D_{\mathbf{q}}D_{\mathbf{p}},
\end{equation}
where $\langle \mathbf{p},\mathbf{q}\rangle=p_2q_1-q_2p_1$ is a symplectic form. As we can see in the above equation, two displacement operators commute if and only if $\langle \mathbf{p},\mathbf{q}\rangle=\langle \mathbf{q},\mathbf{p}\rangle$. Let us now assume that $d$ is a prime number. It is easy to show that
\begin{equation}\label{MUBs_operators}
D_{\mathbf{p}}=\left\{
\begin{array}{c l}
 \tau^{p_2-p_1p_2}(D_{\mathbf{\tilde{p}}})^{p_1} & \mbox{when } p_1\neq0,\\
 (D_{(0,1)})^{p_2} & \mbox{when }p_1=0,
\end{array}
\right.
\end{equation}
where $\mathbf{\tilde{p}}=(1,p_2p_1^{d-2})$ and all operations are modulo $d$.  Notice that the $d^2$ displacement operators can be written as a function of $d+1$ of them. Of course, the functions must be non linear, because all the operators are linearly independent. The eigenvector basis of the displacement operators in prime dimensions are a maximal set of $d+1$ MU bases. \cite{Bandyopadhyay}.

Zauner's conjecture \cite{Zauner1} states that in every dimension $d$ there exists a \emph{fiducial} state $|\phi\rangle$ such that $\{D_{\mathbf{p}}|\phi\rangle\}$ determines a SIC-POVM. That is,
\begin{equation}\label{SIC-POVM1}
    |\langle\phi|D_{\mathbf{p}}|\phi\rangle|^2=\frac{d\delta_{\mathbf{p},\vec{0}}+1}{d+1}~~~\forall\, \mathbf{p}\in\mathbb{Z}_d^2.
\end{equation}
If a pure quantum state $|\phi\rangle$ satisfies the previous condition, then the SIC-POVM is given by $\{D_{\mathbf{p}}|\phi\rangle,\, \mathbf{p}\in\mathbb{Z}_d^2\}$. If such a construction is possible we say that the SIC-POVM is covariant under Weyl-Heisenberg group.

\section{Fiducial states in MU bases decomposition}
Any quantum state $\rho$ acting on a prime dimensional Hilbert space $\mathcal{H}$ can be written as a linear combination of rank-one projectors related to a complete set of $d+1$ mutually unbiased bases. That is,
\begin{equation}\label{MUBs_dec}
    \rho=\sum_{j=0}^d\sum_{k=0}^{d-1}\left(p^j_k-\frac{1}{d+1}\right)\Pi^j_k,
\end{equation}
where
\begin{equation}\label{prob}
    p^j_k=\mathrm{Tr}(\rho\Pi^j_k),
\end{equation}
with $j$ the index of the MU bases family and $k$ the index within the family. We consider that $j=d$ corresponds to the canonical base (eigenvectors of $Z$). Sometimes, this base is denoted with the index $j=\infty$ (see \cite{Appleby1}) and this choice is justified by arguments about the discrete affine plane picture \cite{Gibbons}. Every operator $\Pi^j_k$ is a rank one projector, namely
\begin{equation}\label{vec_MUBs}
    \Pi^j_k=|\varphi^j_k\rangle\langle\varphi^j_k|,
\end{equation}
where $\{|\varphi_k^j\rangle\}$ satisfy the relationship
\begin{equation}
|\langle\varphi^j_k|\varphi^l_r\rangle|^2=\left\{
\begin{array}{c l}
 \frac{1}{d} & \mbox{when } j\neq l,\\
\delta_{k,r} & \mbox{when } j=l,
\end{array}
\right.
\end{equation}
that is, they form a complete set of $d+1$ MU bases. It is also possible to decompose a quantum state as a linear combination of the displacement operators\begin{equation}\label{Desp_dec}
\rho=\sum_{\mathbf{p}\in\mathbb{Z}_d^2}a_{\mathbf{p}}D_{\mathbf{p}},
\end{equation}
where $a_{\mathbf{p}}\in\mathbb{C}$ and $a_{\mathbf{0}}=1/d$. Zauner's conjecture given in Equation (\ref{SIC-POVM1}) can be cast now in the form
\begin{equation}
\label{XXX}
|\mathrm{Tr}(\rho D_{\mathbf{p}})|^2=\frac{d\delta_{\mathbf{p},\vec{0}}+1}{d+1}~~~\forall\, \mathbf{p}\in\mathbb{Z}_d^2.
\end{equation}
Considering Equations (\ref{Desp_dec}) and (\ref{XXX}) and the fact that set $\{D_{\mathbf{p}}\}$ of displacement operators form an orthogonal base we obtain
\begin{equation}\label{coef_disp}
a_{\mathbf{p}}=\left\{
\begin{array}{c l}
 \frac{1}{d} & \mbox{when } \mathbf{p}=\mathbf{0},\\
 \frac{1}{d\sqrt{d+1}}w^{\beta_{\mathbf{p}}} & \mbox{when } \mathbf{p}\neq\mathbf{0}.
\end{array}
\right.
\end{equation}
for a given set of real parameters $\beta_{\mathbf{p}}\in[0,d)$. It is easy to show that the following completness relation holds
\begin{equation}
\sum_{p\in\mathbb{Z}_d^2}\frac{1}{d}D_{\mathbf{p}}\rho D^{\dag}_{\mathbf{p}}=\mathbb{I},
\label{YYY}
\end{equation}
for any set $\{\beta_{\mathbf{p}}\}$. In order to deduce a relationship between the set of coefficients $\{p^j_k\}$ given by Equation (\ref{MUBs_dec}) and the set of coefficients $\{a_{\mathbf{p}}\}$ given by Equation (\ref{Desp_dec}) we need an expression of the displacement operators as a function of the MU bases projectors. Taking into account that every operator $D_{\mathbf{p}}$ has the same set of eigenvalues $\{\omega^k\}$ (with $k=0,\dots,d-1$) we obtain
\begin{equation}\label{decomp1}
D_{\mathbf{p}}=\left\{
\begin{array}{c l}
 \tau^{p_2-p_1p_2}\sum_{k=0}^{d-1}\omega^{kp_1}\Pi^{\tilde{p}_2}_k & \mbox{when } p_1\neq0,\\
\sum_{k=0}^{d-1}\omega^{kp_2}\Pi^{0}_k& \mbox{when }p_1=0.
\end{array}
\right.
\end{equation}
Putting Equations (\ref{decomp1}) into Equation (\ref{Desp_dec}) and considering Equation (\ref{coef_disp}) we have
\begin{eqnarray}
\rho&=&\sum_{p_2=1}^{d}\sum_{k=0}^{d-1}\frac{1}{d(d+1)}\Pi^{p_2}_k
 \nonumber\\&+&\sum_{p_2=1}^{d}\sum_{k=0}^{d-1}\sum_{p_1=1}^{d-1}a_{(p_1,p_1p_2)}\tau^{p_1p_2-p_1^2p_2}\omega^{kp_1}\Pi^{p_2}_k
			 \nonumber\\&+&\sum_{k=0}^{d-1}\left(\frac{1}{d(d+1)}+\sum_{p_2=1}^{d-1}a_p\omega^{kp_2}\right)\Pi^0_k.
			\label{FIDUCIAL-MUBS}
\end{eqnarray}
From the last result and considering Equation (\ref{coef_disp}), we have the following relationship
\begin{equation}\label{decomp2}
p^j_k=\frac{1}{d}+\frac{1}{d\sqrt{d+1}}\sum_{r=1}^{d-1}\omega^{\alpha^j_r+kr},
\end{equation}
for every $j=0,\dots,d$ and $k=0,\dots,d-1$, where the $d^2-1$ phases have the form
\begin{equation}
\omega^{\alpha^j_r}=\left\{
\begin{array}{c l}
 \omega^{\beta_{(r,jr)}}\tau^{jr-jr^2} & \mbox{when } j\neq0,\\
\omega^{\beta_{(0,r)}} & \mbox{when }j=0,
\end{array}
\right.
\end{equation}
Given that every probability $p^j_k$ is a real number and considering that $d$ is an odd prime number we find the symmetry
\begin{equation}\label{restrictions1}
    \alpha^j_r=-\alpha^j_{d-r}.
\end{equation}
for every $j=0,\dots,d$ and $r=1,\dots,d-1$. Therefore, Equation (\ref{decomp2}) is reduced to
\begin{equation}\label{solution}
    p^j_k=\frac{1}{d}+\frac{2}{d}\sqrt{\frac{1}{d+1}}\sum_{r=1}^{(d-1)/2}\cos(\alpha^j_r+2\pi kr/d).
\end{equation}
This characterization of the probability distributions for fiducial operators in MU bases decomposition is our first result. Let us note that the normalization conditions are implicit in the last equation. Additional non-trivial restrictions to the phases $\{\alpha^j_r\}$ can be obtained from considering $0\leq p^j_k$ on Equation (\ref{solution}), that is
\begin{equation}\label{restrictions2}
-\frac{\sqrt{d+1}}{2}\leq\sum_{r=1}^{(d-1)/2}\cos(\alpha^j_r+kr),
\end{equation}
whereas $p^j_k\leq1$ gives us a trivial upper bound. All operators $\rho$ given by Equation (\ref{MUBs_dec}) with probability distributions $\{p^j_k\}$ given by Equation (\ref{solution}) build the manifold $\Omega$. Any operator in $\Omega$ can be used to construct a set of $d^2$ operators satisfying Equations. (\ref{XXX}) and (\ref{YYY}). However, we cannot guarantee the existence of quantum states in $\Omega$, that is, positive semidefinite operators of unitary trace. Indeed, a fiducial state $\rho$ expressed in Equation (\ref{MUBs_dec}) must be pure. An adequate set of phases $\{\alpha^j_r\}$ such that $\rho\geq0$ and $\rho\in\Omega$ give us a fiducial pure state. Another way to show this is by considering the following proposition \cite{Appleby2}:
\begin{prop}\label{propApp}
Let $|\phi\rangle$ a quantum state belonging to a $d-$dimensional Hilbert space with $d$ prime and let $\{p^j_k\}$ the set of $d+1$ probabilities distributions associated to $|\phi\rangle$. Then, $|\phi\rangle$ is fiducial if and only if the following conditions hold
\begin{enumerate}
\item $\sum_{k=0}^{d-1}(p^j_k)^2=\frac{2}{d+1}$, $\forall~j=0,\dots,d$.
\item $\sum_{k=0}^{d-1}p^j_kp^j_{k+r}=\frac{1}{d+1}$, $\forall~r=1,\dots,d-1$.
\end{enumerate}
\end{prop}
It is easy to show that any $\rho\in\Omega$ satisfies the conditions \emph{(i)} and \emph{(ii)} for every prime dimension. However, we cannot prove that $\Omega$ contain pure states. This is the reason why we do not prove the existence of fiducial states in prime dimensions.

\section{Purity conditions}
Any operator contained in the set $\Omega$ is hermitian but not necessarily a quantum state. Our interest here is to find a reduced number of conditions over the phases $\{\alpha^j_r\}$ to single out in $\Omega$ a pure quantum state. The following proposition helps to this end \cite{Jones}:
\begin{prop}\label{prop1}
Let $\rho$ be an hermitian operator, and suppose that\\ $\mathrm{Tr}(\rho^2)=\mathrm{Tr}(\rho^3)=1$. Then, $\rho$ is a rank-one projector.
\end{prop}
Let us note that if $\mathrm{Tr}(\rho)=1$ and $\mathrm{Tr}(\rho^2)=1$ the only way to have a positive semidefinite  operator $\rho$ is that $\mathrm{Tr}(\rho^3)=1$. If this does not hold, then $\rho$ has one negative eigenvalue, at least. This fact is easy to understand because the only quantum states having $\mathrm{Tr}(\rho^2)=1$ are the pure states. In our case, $\mathrm{Tr}(\rho)=1$ is implicit in the MUBs decomposition given in Equation (\ref{MUBs_dec}). We can also easily deduce from Equation (\ref{decomp2}) that
\begin{equation}
  \sum_{j,k}(p^j_k)^2=2,
\end{equation}
for every value of the phases $\{e^{i\alpha^j_r}\}$. On the other hand, it is known a general property relating probabilities in MUBs decomposition with the purity of a quantum state \cite{Larsen}, namely:
\begin{equation}\label{purity}
    \sum_{j,k}(p^j_k)^2=\mathrm{Tr}(\rho^2)+1.
\end{equation}
Combining the last two equations we obtain
\begin{equation}
\mathrm{Tr}(\rho^2)=1.
\end{equation}
Thus, condition \emph{1)} of Proposition \ref{prop1} is implicit in the probability distributions $\{p^j_k\}$ given by Equation (\ref{solution}). Therefore, we only need to impose the condition
\begin{equation}\label{tr3}
\mathrm{Tr}(\rho^3)=1,
\end{equation}
in order to have a pure state in $\Omega$. Taking into account Equation (\ref{tr3}) and the MU bases decomposition given in Equation (\ref{FIDUCIAL-MUBS}) we obtain
\begin{equation}
    \sum_{{j\atop k}{l\atop m}{r\atop s}}p^j_k\,p^l_m\,p^r_s\,\mathrm{Tr}\,(\Pi^j_k\,\Pi^l_m\,\Pi^r_s)=
    \mathrm{Tr}\,(\rho^3)+d+6,
    \label{PURITY-CONDITION}
\end{equation}
which must be satisfied by the probability distributions $\{p^j_k\}$ (depending on $\{\alpha^j_k\}$) to guarantee that they represent a physically admissible fiducial pure state. Equation (\ref{PURITY-CONDITION}) indicates that the problem of the existence of  Weyl-Heisenberg covariant SIC-POVMs is equivalent to demonstrate that the function
\begin{equation}\label{problem}
    F(\{p^j_k\})=\sum_{{j\atop k}{l\atop m}{r\atop s}}p^j_k\,p^l_m\,p^r_s\,\mathrm{Tr}\,(\Pi^j_k\,\Pi^l_m\,\Pi^r_s)
\end{equation}
achieves a maximum value of $d+7$ for some $\rho$ in $\Omega$. We can affirm that the maximum value belongs to the interval $[d+5,d+7]$, but so far we have not been able to proof that $F=d+7$ is reached for every prime dimension $d$. Numerical evidence tells us that this maximal number is reached in $\Omega$ in every dimension $d\leq67$ \cite{Grassl5} and, consequently, we have a strong evidence of its existence in every prime dimension.

Let us establish an upper bound for the probabilities $p^j_k$. From Equation (\ref{solution}) we can bound the cosines, obtaining
\begin{equation}
    p^j_k\leq\frac{1}{d}+\frac{d-1}{d}\sqrt{\frac{1}{d+1}},
\label{UPPERBOUND}
\end{equation}
for every $k=0,\dots,d-1$ and $j=0,\dots,d$. This bound is highly non-trivial and it may be useful to reduce computational time in order to find numerical solutions in higher dimensions. Another interesting property is that any $\rho\in\Omega$ minimizes the total quadratic R\'enyi entropy
\begin{equation}
T=\sum_{j=0}^d H_j=-\sum_{j=0}^d\log_2\left(\sum_{k=0}^{d-1}(p^j_k)^2\right).
\end{equation}
Considering Equation (\ref{decomp2}) we obtain the minimal quadratic R\'enyi entropy as a function of the dimension $d$ only, that is
\begin{equation}
T=(d+1)\log_2\left(\frac{d+1}{2}\right).
\end{equation}
This result is a consequence of the convexity of the logarithm, Jensen's inequality and Equation (\ref{decomp2}). The total quadratic R\'enyi entropy is useful to define an entropic uncertainty principle, as we can see in \cite{Appleby2}. In the same work, it was shown that fiducial states in prime dimensions are minimum uncertainty states if the total uncertainty $T=\sum_{j=0}^{d-1}H_j$ is considered. Here, we have found that any hermitian operator $\rho\in\Omega$ minimizes this uncertainty principle.

\section{Summary}
We have studied the problem of constructing Weyl-Heisenberg covariant SIC-POVMs in prime dimensions. Considering useful properties of the displacement operators we have constructed a $(d^2-1)/2-$dimensional manifold $\Omega$ of hermitian operators $\rho$ that contains the complete set of Weyl-Heisenberg covariant fiducial operators. Any $\rho\in\Omega$ satisfy $\mathrm{Tr}(\rho)=1$ and $\mathrm{Tr}(\rho^2)=1$. The existence of fiducial pure states $\rho\in\Omega$ is equivalent to show that the global maximum $\mathrm{Tr}(\rho^3)=1$ is attainable in every dimension for some $\rho\in\Omega$. Finally, we also deduced an upper bound for each probability of a fiducial operator in MU bases decomposition in prime dimension and proved that any operator $\rho\in\Omega$ minimizes an entropic uncertainty principle. This allow us to think fiducial states as \emph{coherent states} in prime dimensions \cite{Appleby2}. We would like to conclude by noting a connection between fiducial states and entanglement: The matrix  operators in $\Omega$ can be map onto a set $\tilde\Omega$ in a $d^2$-dimensional product space $L$ by concatenating one row after the other. This lineal space $L$ can be decomposed as $L_1\otimes L_2$, where both $L_1$ and $L_2$ are $d$-dimensional and elements in the canonical base of $L$ are identified with tensor products of elements in the canonical bases of $L_1$ and $L_2$. It turns out that an operator in $\Omega$ is a fiducial pure state if and only if it corresponds to a separable vector of $\tilde\Omega$ in $L_1\otimes L_2$. We hope our results, together with very well known properties of the pure/separable states manifold, help us to proof the existence of SIC-POVMs in every prime dimension without finding explicit expressions for them.

\section{Acknowledgments}

This work was supported by Grants CONICyT PFB-0824, MSI P10-030-F, and FONDECyT  Grants N$^{\text{\underline{o}}}$ 1080383 and N$^{\text{\underline{o}}}$ 3120066. R. S. acknowledges support from CONICyT.

\bibliographystyle{elsarticle-num}

\end{document}